# Superconducting quantum oscillations and anomalous negative magnetoresistance in a honeycomb nano-patterned oxide interface superconductor

Yishuai Wang[1], Siyuan Hong[1], Wenze Pan[1], Yi Zhou[2,3], and Yanwu Xie[1,4,5,6]*

[1]School of Physics, and State Key Laboratory for Extreme Photonics and Instrumentation, Zhejiang University, Hangzhou 310027, China
[2]Beijing National Laboratory for Condensed Matter Physics & Institute of Physics, Chinese Academy of Sciences; Beijing 100190, China.
[3]Kavli Institute for Theoretical Sciences and CAS Center for Excellence in Topological Quantum Computation, University of Chinese Academy of Sciences; Beijing 100190, China.
[4]College of Optical Science and Engineering, Zhejiang University, Hangzhou 310027, China
[5]Hefei National Laboratory; Hefei 230088, China.
[6]Collaborative Innovation Center of Advanced Microstructures, Nanjing University; Nanjing 210093, China

*To whom correspondence should be addressed. E-mail: ywxie@zju.edu.cn

**Abstract**

The extremely low superfluid density and unprecedented tunability of oxide interface superconductors provide an ideal platform for studying fluctuations in two-dimensional superconductors. In this work, we have fabricated a $LaAlO_3/KTaO_3$ interface superconductor patterned with a nanohoneycomb array of insulating islands. Little-Parks-like magnetoresistance oscillations have been observed, which are dictated by the superconducting flux quantum $h/2e$. Moreover, an anomalous negative magnetoresistance (ANMR) appears under a weak magnetic field, suggesting magnetic-field-enhanced superconductivity. By examining their dependences on temperature, measurement current, and electrical gating, we conclude that both phenomena are associated with superconducting order parameter: The $h/2e$ oscillations provide direct evidence of Cooper pair transport; the ANMR is interpreted as a consequence of multiple connected narrow superconducting paths with strong fluctuations.

The superconductivity is governed by the formation and the phase coherence of Cooper pairs. In most conventional superconductors, the energy associated with destroying the phase coherence is larger than the Cooper pair-breaking energy, and thus the phase is always coherent once Cooper pairs are formed [1]. However, the situation becomes different in systems with a low superfluid density [1,2], notably the underdoped cuprate high-temperature superconductors [3] and the conventional superconductors with strong disorders and low dimensionality [4]. In these systems the phase may be incoherent even after Cooper pairs are formed [1]. Furthermore, the unavoidable disorders in real samples may cause emergent granularity of the superconductivity [2]. These phase and spatial fluctuations are believed to account for many well recognized phenomena including pre-formed Cooper pairs [5], anomalous metallic state [2], and superconductor-insulator quantum phase transition [2].

The two-dimensional oxide interface superconductors (2DOISs), represented by $LaAlO_3/SrTiO_3$ (LAO/STO) [6] and $LaAlO_3$ (or EuO)/$KTaO_3$ (LAO/KTO) [7–9], are of extremely low superfluid density [10–12]. Their corresponding three-dimensional charge carrier density is $\sim 10^{19}$-$10^{20}$ cm$^{-3}$, which is about 2-3 (or 1-2) orders of magnitude less than that of most conventional superconductors (or cuprate high-temperature superconductors). Moreover, only a small fraction of their charge carriers participate in the superconductivity [10,11]. Therefore, phase fluctuations are expected to be strong in 2DOISs, which is experimentally supported by the presence of preformed Cooper pairs in LAO/STO nanowires [13] and the fragileness of zero-resistance state in KTO interface superconductors [8,11]. Additionally, STO- and KTO-based oxide interfaces are particularly intriguing due to their broken inversion symmetry and strong spin-orbit coupling [7,14–22]. On the other hand, 2DOISs have excellent tunable abilities. Their ground states can be tuned by simply applying an electric field across STO [23] or KTO [8] substrate, via modulation in carrier density [23] or effective disorder [8]. Even more intriguingly, 2DOISs can be controlled locally and nonvolatilely using a unique technique called conductive atomic force microscope (cAFM) lithography [12,24–27]. The extremely low superfluid density and the unprecedented tunability make 2DOISs an intriguing platform in exploring the rich physics involving low-dimensional superconducting fluctuations.

In this work, we describe transport measurements on a LAO/KTO 2DOIS patterned with a nanohoneycomb array of insulating islands. We observe two spectacular quantum phenomena: the Little-Parks-like magnetoresistance oscillations dictated by the superconducting flux quantum $h/2e$ [28–32] and an anomalous negative magnetoresistance (ANMR) under a weak magnetic field. The $h/2e$ oscillation is a key signature of the superconducting electron pairing, but it has only been observed in few cases in 2DOISs [12,33], with experimentally measured periods deviating from the expected superconducting flux quantum yet. ANMR is a striking but not-well-understood quantum phenomenon that has been observed mostly in one-dimensional superconducting systems [34–40]. The observation of both phenomena in the same LAO/KTO device not only provides a solid evidence of Cooper pair transport in 2DOISs, but also demonstrates an excellent example that captures the importance of superconducting fluctuations in a tunable manner.

**Device structure**

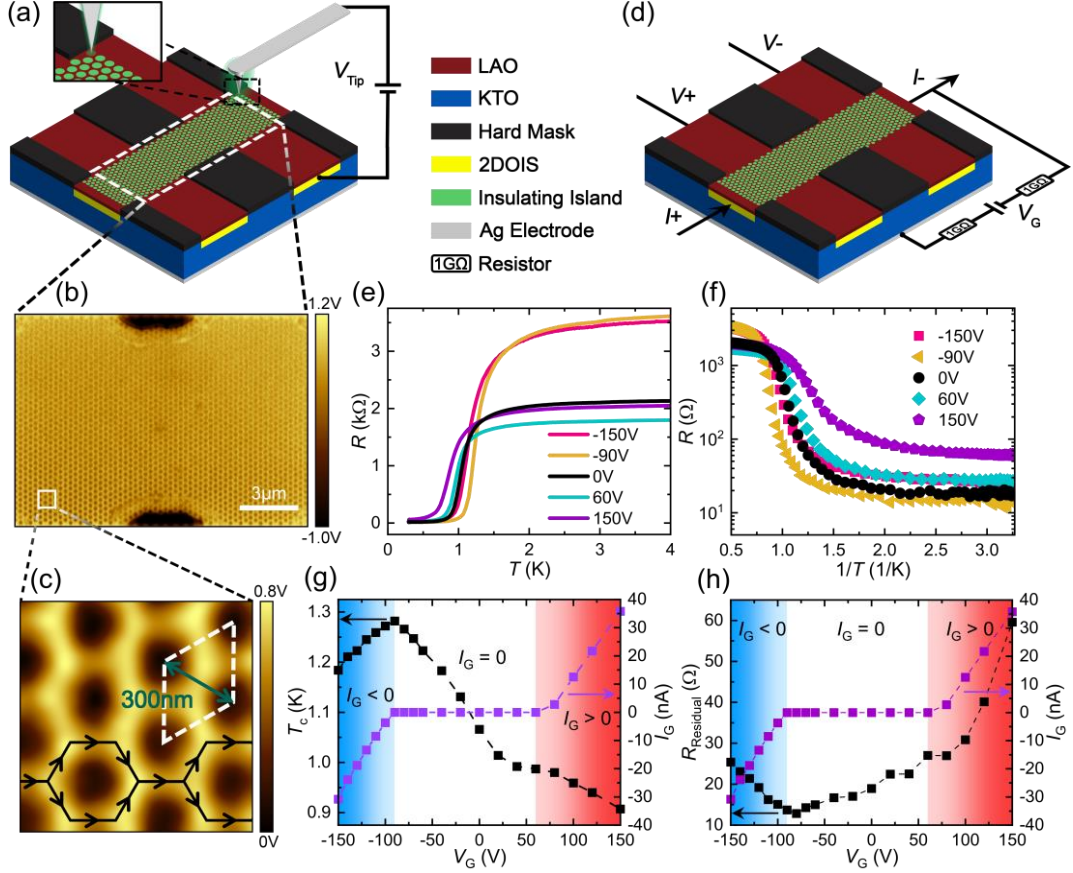

FIG. 1. Device structure and transport characterization. (a) Illustration of creating a nanohoneycomb array of insulating islands by cAFM lithography on the central channel of a Hall-bar shaped LAO/KTO 2DOIS. (b) Surface potential image of the patterned area as indicated by the dashed white lines in (a). The surface morphology image measured simultaneously is shown in Fig. S2. (c) An enlarged view of one part of (b). The dark dots indicate the insulating islands; the bright mesh is the remaining superconducting paths. The dashed diamond indicates one unit cell of the array. The arrowed lines in the bottom schematically show a set of current paths that surround insulating islands. (d) The measurement configuration of the patterned 2DOIS device. (e) Temperature dependence of resistance at representative $V_G$ values. (f) Arrhenius plot of the same $R(T)$ curves in (e). The $V_G$ dependences of $I_G$, (g) the mid-point $T_c$ and (h) the residual resistance $R_{Residual}$. The "$I_G$ = 0" means that the monitored $I_G$ value is less than 0.1 nA.

The LAO/KTO 2DOIS is formed by depositing a 6-nm-thick LAO film on a KTO(110) single-crystalline substrate that has been pre-patterned into a Hall bar configuration with a 10-μm-wide central bridge. A photograph of the device is shown in Fig. S1 [41]. The creation of a nanohoneycomb array of electrically insulating islands is as schemed in Fig. 1(a), using cAFM lithography. These islands can be well detected in surface potential images (seen as dark dots in Figs. 1(b) and 1(c)) which were captured using the same cAFM probe in the non-contact mode. The distance between the centers of two adjacent islands, which is the sketched lattice constant, is 300 nm, and the area of one unit cell is around 78000 nm$^2$ (Fig. 1(c)). The transport measurement configuration is shown in Fig. 1(d). A direct current (DC) method is used. During measurement, a gate voltage ($V_G$) is applied across KTO, and two 1-GΩ resistors are used to protect the device from electrical breakdown as well as to control the magnitude of the out-of-plane leakage current ($I_G$). More details can be found in Supplementary Material (SM) [41].

**Transport characterization**

Figure 1(e) shows representative curves of temperature-dependent resistance at different $V_G$ values (the full set of curves is shown in Fig. S4(a)). With decreasing temperature, a superconducting transition occurs in all the curves, and the optimal mid-point $T_c$ can reach 1.28 K at $V_G = -90$ V (or ~1.05 K at $V_G = 0$ V) (Fig. 1(g)). However, as demonstrated in Fig. 1(f) [the Arrhenius plot of the same $R$(T) curves in Fig. 1(e)], a true superconducting ground state ($R = 0$) is absent, and a residual resistance ($R_{Residual}$) that is about $3\times10^{-3}$ to $4\times10^{-2}$ times the corresponding normal-state value is observed, indicating an anomalous metallic state. For $V_G$ between -90 V and 60 V, the $I_G$ is negligible (the $I_G = 0$ region), and the $V_G$ dependence of transport properties (see Fig. S4(b)) is similar to previous gating studies (for example, see Fig. S5) [8,15,21]. However, at large $V_G$ values ($V_G < -90$ or $> 60$ V), a non-negligible $I_G$ is present (Figs. 1(g) and 1(h)), which lowers the effective $V_G$ (saturating roughly for $V_G < -95$ or $> 70$ V; see Fig. S6). With increasing $I_G$, $T_c$ decreases (Fig. 1(g)) and $R_{residual}$ increases (Fig. 1(h)), indicating the presence of stronger superconducting fluctuations. As will be shown below, $I_G$ has a profound effect on the around-zero-field magnetoresistance.

**Magnetoresistance oscillations**

We first show that the device exhibits Little-Parks-like magnetoresistance (MR) oscillations. Figure 2(a) plots the magnetoresistance, $R(H)$, of the device measured at $V_G = 0$ V, $T = 0.3$ K, and $I_{DC} = 350$ nA (a curve with magnetic field up to 3 T is shown in Fig. S8(b)). The $R(H)$ oscillates with a well-defined period of 26.5 mT. This period is in good agreement with one superconducting flux quantum $\Phi_0 = h/2e$ threading an area of one unit cell of the patterned array, where $h$ is the Planck constant and $e$ is the elementary charge. The $R(H)$ minima occur when $\Phi = n\Phi_0$. Similar MR oscillations with the period of $\Phi_0/S$ were also observed in devices with larger sketched lattice constants (450 nm and 900 nm, see Figs. S15 and S16). The oscillations are still detectable up to $T = 1.0$ K, which is almost the $T_c$ for $V_G = 0$ V (Figs. 2(b) and 2(c)). The period of oscillations is independent of the measurement temperature (Figs. 2(b) and 2(c)) and $V_G$ (Fig. 2(e)). In addition to the magnetoresistance oscillations, as shown in Fig. 2(c), the temperature contour of constant resistance (indicated by the white boundary between the red and blue regions) also oscillates with $\Phi/\Phi_0$, with maxima occurring when $\Phi = n\Phi_0$, similar to the critical temperature oscillations observed in the Little-Parks effect [29,42]. Clearly, the presently observed Little-Parks-like charge $2e$ MR oscillation is a manifestation of the collective quantum behavior of superconducting pairs passing the patterned 2DOIS.

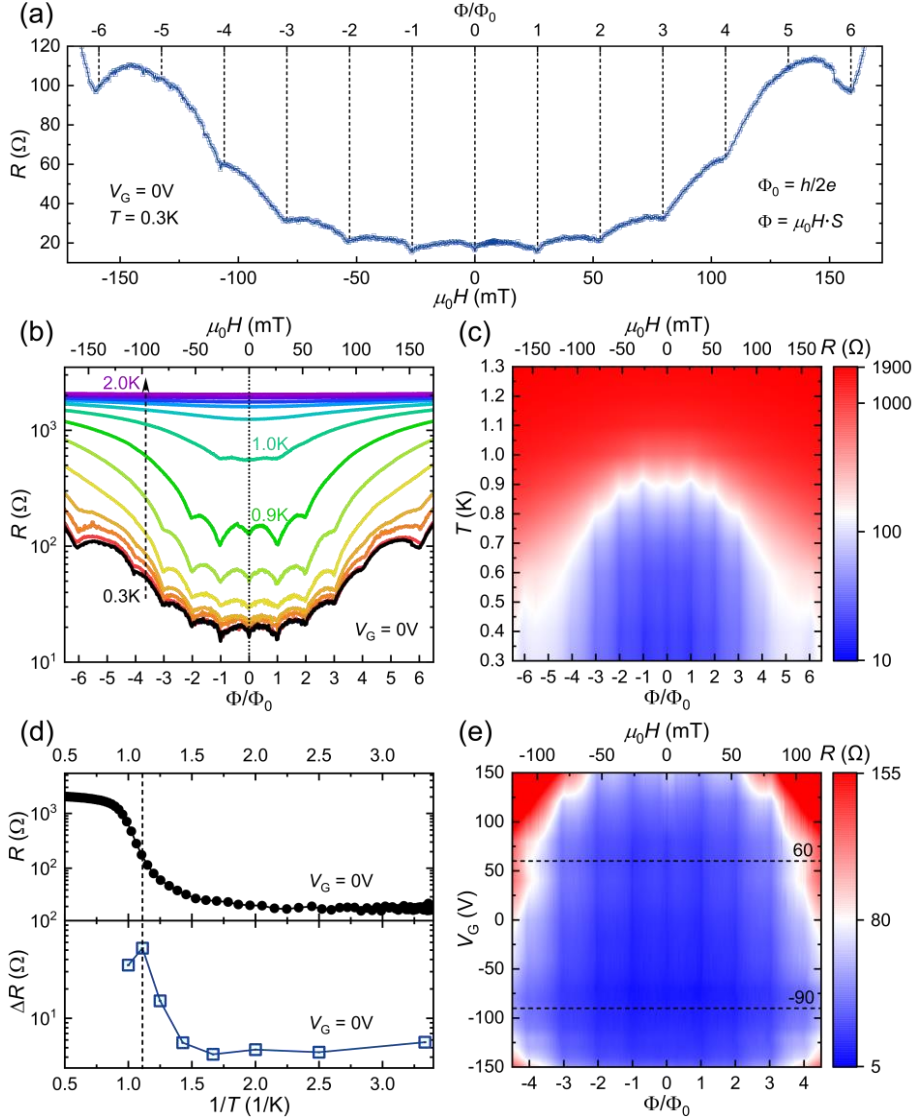

FIG. 2. Little-Parks-like effect of the patterned 2DOIS device measured at $I_{DC}$ = 350 nA. (a) Oscillations of the device resistance as a function of the perpendicularly applied magnetic field. The top-axis shows the magnetic flux for the device geometry, $\Phi = \mu_0 HS$, where $S$ is the area of one unit cell (see Fig. 1(c)), normalized by a superconducting flux quantum $\Phi_0 = h/2e$. (b) $R$ versus $H$ (or $\Phi/\Phi_0$) at different temperatures and (c) color contour plot of $R$ versus $H$ (or $\Phi/\Phi_0$) and $T$, which displays temperature-independent Little-Parks-like oscillations. (d) The saturated behavior of the oscillations in the anomalous metallic state. Upper panel: Temperature dependence of resistance at zero magnetic field (same data as in Fig. 1(f)). Lower panel: Temperature dependence of the amplitude of the first Little-Parks-like oscillation. (e) Color contour plot of $R$ versus $H$ (or $\Phi/\Phi_0$) and $V_G$, which displays $V_G$-independent periodic oscillations. The line graph is shown in Fig. S9.

We note that superconducting oscillations have been demonstrated in superconducting quantum interference devices (SQUIDs) constructed with LAO/STO [33] and LAO/KTO [12]. However, in those 2DOIS-based SQUIDs, the experimentally measured period in the magnetic field was several times smaller than the expected superconducting flux quantum through the designed loop. This deviation was attributed to the Meissner effect [33] or the mesoscopic dimension of the superconducting leads [12] in the devices. By contrast, in our present devices, the period of the oscillations is strictly consistent with the $h/2e$ superconducting flux quantum through one unit cell (Fig. S16), which unambiguously demonstrates the existence of charge-$2e$ Cooper pairs in the KTO-based 2DOIS.

In the original Little-Parks effect [29], the MR oscillations reflect periodic changes in $T_c$, and thus only occur in the vicinity of $T_c$. By contrast, the presently observed MR oscillations survive down to 0.3 K (the lowest temperature we have used, much lower than $T_c$) (Fig. 2(d)), and have amplitudes much larger than $\Delta R(H) = \Delta T_c(dR/dT)$ (see SM and Figs. S8(c) and S8(d)). The large amplitude of MR oscillations in periodic superconducting networks can be described by the modulation of thermally activated [31,43,44] or current-excited [45] vortex motion. For KTO-based 2DOIS, the effect of vortices on transport properties should be significant due to its relatively strong disorder [8] and intrinsic 2D nature [7–9]. Previous work has proposed that the vortex plays a central role in the intermediate disorder regime [46]. Due to the extremely low superfluid density of LAO/KTO 2DOIS [11,12], its Meissner effect, which is crucial in the current-excited moving vortices scenario [45], is expected to be weak. Thus, we introduce the thermally activated fluxoid dynamics model [31,43,44] to analyze the MR oscillations quantitatively. As shown in Figs. S8(c) and S8(d), our data fit well in the relatively high-temperature regime ($T > 0.7$ K), which corresponds to the superconducting transition range, supporting the thermally activated vortex scenario [31,43,44]. For devices with a true zero-resistance superconducting state [31,43,44], vortices are expelled out or frozen below the superconducting transition temperature, leading to the vanishing of MR oscillations. However, this is not the case for our device, as it shows an anomalous metallic state below that temperature (Figs. 1(f) and 2(d)). Correspondingly, a non-vanishing and saturated MR oscillation persists down to the lowest temperature (Fig. 2(d)).

We recall that saturated MR oscillations persisting to the lowest temperatures have been observed in amorphous Bi [28], TiN [47], and $YBa_2Cu_3O_{7-x}$ [48,49] films patterned with an ordered array of holes. These observations were understood within the framework of Cooper pair quantum interference effects in Josephson junction arrays (JJAs) [28,47–50]. Clearly, our devices are equivalent to those nanopatterned superconductors, except that holes are replaced with insulating islands (Figs. 1(b) and 1(c)). Moreover, the local superconducting channel width revealed by the surface potential image (~100 nm, see (Fig. 1(c)) is comparable to the zero-temperature superconducting coherence length (~20 nm, see Fig. S8(b)), and is sufficiently narrow for the weak link to behave as a proximity-effect-generated Josephson junction [47]. Therefore, a similar JJA scenario is expected to apply. As shown in Fig. S10, by subtracting a smooth background from the $R(H)$ curve, the MR oscillations were extracted and well-fitted by $\Delta R \propto |\sin(\pi\Phi/\Phi_0)|$, consistent with the JJA scenario [51]. Following this scenario, the observed anomalous metallic state can be interpreted as either the quantum tunneling of vortices in JJAs with Ohmic dissipation [50] or the localization of Cooper pairs in JJAs [48]. We note that an anomalous metallic state was also observed in the non-patterned uniform LAO/KTO interfaces [8], where strong fluctuations may lead to emergent superconducting granularity [2], forming effective JJAs.

**Anomalous negative magnetoresistance (ANMR)**

Having established the Little-Parks-like magnetoresistance oscillation, we turn our attention to the ANMR that can coexist with the oscillations. As shown in Fig. 3(b), a remarkable resistance peak around zero magnetic field is observed when both the $V_G$ and the measurement current $I_{DC}$ are large. This resistance peak gives a counterintuitive ANMR in which a magnetic field enhances, rather than suppresses, superconductivity.

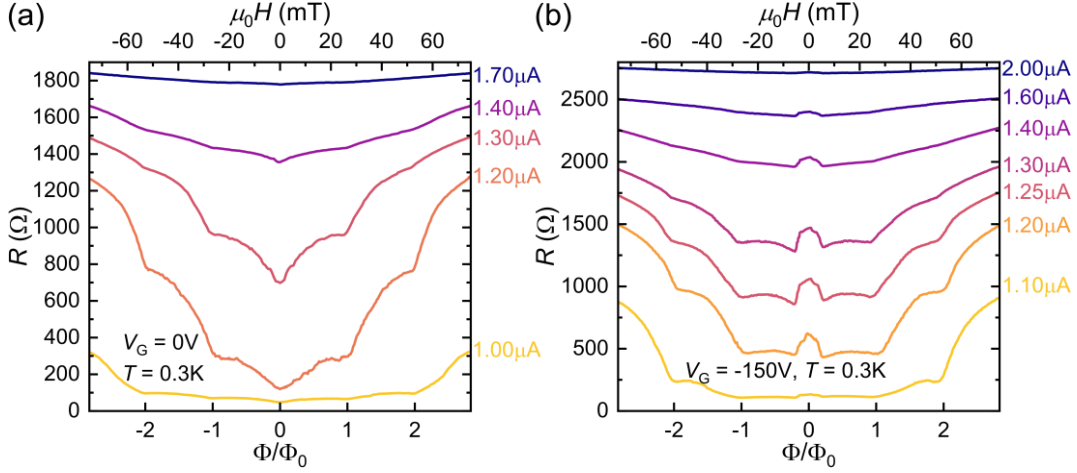

FIG. 3. $R(H)$ curves measured at different DC currents. (a) For $V_G = 0$ V (b) For $V_G = -150$ V. A remarkable resistance peak around $H = 0$ is observed in (b).

To gain further insights of this unexpected ANMR, we examine its dependences on $I_{DC}$, temperature, and $V_G$. In order to make a quantitative comparison, we define the magnitude of ANMR as $\Delta R_{peak}$ (as indicated in Fig. 4(a)). Three distinct observations are observed. First, for a given temperature and $V_G$, the ANMR is strongest when $I_{DC}$ is around the superconducting critical current $I_c$, (Figs. 4(b), 4(c), 4(f), S12, and S13). A close examination of the current-voltage ($IV$) curves reveals that it originates from the field-induced increase in $I_c$ (Fig. S14). Second, the ANMR is strongest at the lowest temperature (0.3 K) and becomes weaker gradually with increasing temperature until $T_c$, where it fully vanishes (Figs. 4(d) and 4(e)). This observation is a strong indication that the involved processes are quantum rather than thermal. Third, the ANMR is accompanied by the presence of an out-of-plane $I_G$, and becomes stronger with increasing the absolute value of $I_G$ (Figs. 4(g)-4(i)). Based on the second observation we know that the major effect of $I_G$ cannot be Joule heating, otherwise the ANMR would be stronger at smaller, rather than larger, $I_G$ values. In addition, the stray current of $I_G$ can be excluded as the origin of ANMR because $I_G$ is 1-2 orders smaller than $I_{DC}$. Furthermore, if the stray current were the origin, we would expect that the ANMR is most prominent when $I_{DC}$ is small, which is contrary to the experimental observations (Figs. 4(b)-4(d)). Weak localization may also be excluded as the possible mechanism because its negative MR is typically 2-3 orders of magnitude smaller than the presently observed ANMR, and should be irrelevant with superconductivity [52,34,35]. In addition, we note that an interesting negative MR was once observed in LAO/STO 2DOIS [53]. However, the presently observed ANMR is apparently different from it because that one appears in a parallel (vs perpendicular) magnetic field, near $T_c$ (vs strongest at a temperature far below $T_c$), and the samples were uniformly large (vs nanopatterned).

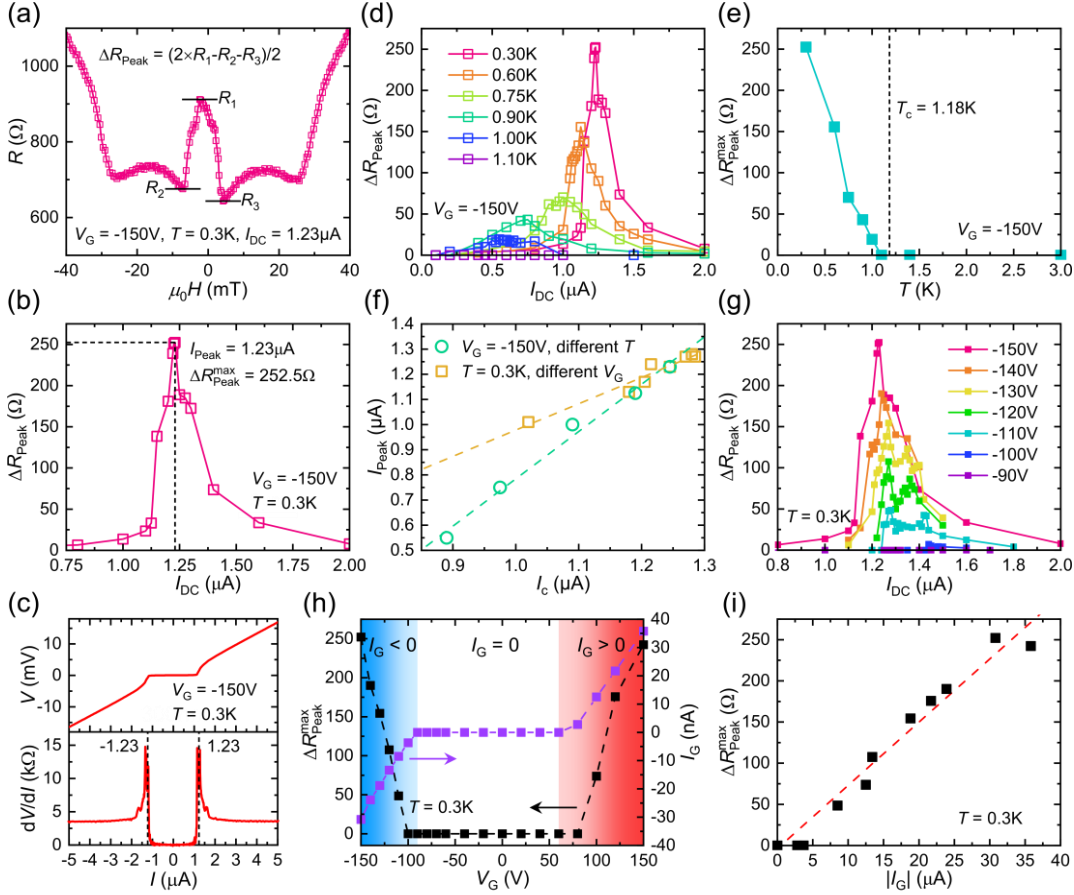

FIG. 4. ANMR. (a) A representative $R(H)$ curve showing ANMR. $\Delta R_{\text{Peak}}$ defines the amplitude of ANMR. (b) Dependence of $\Delta R_{\text{Peak}}$ on $I_{\text{DC}}$ for $V_G$ = -150 V and $T$ = 0.3 K. $I_{\text{Peak}}$ and $\Delta R_{\text{Peak}}^{\max}$ are defined as labeled. The same definition of $\Delta R_{\text{Peak}}^{\max}$ is used in the following (Figs. 4(e), 4(h), and 4(i)). (c) Upper panel: Current-voltage (IV) relation for $V_G$ = -150 V and $T$ = 0.3 K. Lower panel: The peaks in differential resistance, $dV/dI$, defines the critical current $I_c$. The dashed lines indicate the position of $I_{\text{Peak}}$. (d) Dependence of $\Delta R_{\text{Peak}}$ on $I_{\text{DC}}$ at different temperatures, for $V_G$ = -150 V. (e) The temperature dependence of $\Delta R_{\text{Peak}}^{\max}$ for $V_G$ = -150 V. The corresponding $R(H)$ curves are shown in Fig. S11. (f) $I_{\text{Peak}}$ versus $I_c$. Squares: measured at $T$ = 0.3 K and different $V_G$ values; Circles: measured at $V_G$ = -150 V and different temperatures. (g) Dependence of $\Delta R_{\text{Peak}}$ on $I_{\text{DC}}$ at different $V_G$ values. (h) The $V_G$ dependence of $\Delta R_{\text{Peak}}^{\max}$ versus the $V_G$ dependence of $I_G$. (i) $\Delta R_{\text{Peak}}^{\max}$ versus the absolute value of $I_G$.

Previously, a similar ANMR has been observed in 1D superconducting systems with strong fluctuations [34–40], although a consensus on its detailed physical mechanisms is still lacking [34–37,54–56]. We attribute the currently observed ANMR to a similar scenario, as our device can be viewed as a mesh of multiple connected 1D superconducting paths. In examining this scenario, we also created ~100-nm-wide single nanowires of LAO/KTO 2DOIS, and although weak, ANMR was indeed observed (see Fig. S17). Additionally, after subtracting the MR oscillations shown in Fig. 2(a), a weak ANMR is still visible (see Fig. S10). The observation of ANMR without the presence of $I_G$ in both cases suggests that the role of $I_G$ is to enhance ANMR, rather than being indispensable for it. In this context, the striking enhancement of ANMR in our device can be explained by that $I_G$, as a dissipation source, enhances superconducting fluctuations (as already indicated in Figs. 1(g) and 1(h)).

In light of previous studies on 1D superconducting systems [34–37,39,54,56], we consider two likely mechanisms to explain the underlying physical processes. The first mechanism is the nonequilibrium charge imbalance process at normal-superconducting (N-S) boundaries [34,56], where the magnetic field reduces the charge imbalance relaxation time and consequently, the N-S boundary resistance. The second mechanism, proposed by Kivelson and Spivak [57,58], suggests that strong fluctuations cause the Josephson coupling between superconducting domains or in superconducting-normal-superconducting (S-N-S) junctions to become random in magnitude and sign, resulting in ANMR at weak magnetic fields. The former was applied to explain the ANMR in systems such as Al [34] and InO [35] wires, while the latter was used to explain the ANMR in Pb wires [36]. These two mechanisms are not mutually exclusive, as the existence of the N-S boundary and strong fluctuations are central to both. Regarding the present ANMR, most observations mentioned above can be reasonably explained by both mechanisms, except for the dependence on $I_{DC}$. Santhanam *et al.* [34] suggested that, in the nonequilibrium charge imbalance mechanism, the amplitude of ANMR decreases with $I_{DC}$, which contradicts our observations for $I_{DC} < I_c$. In this regard, the second mechanism appears to be more consistent with our findings. We also note that Xiong *et al.* [36] observed that ANMR emerged in the zero-current limit, while in our case, the ANMR is stronger near $I_c$. This discrepancy may arise from variations in superconducting fluctuations across different systems. Further studies are needed to fully elucidate the underlying physical mechanisms.

**Summary**

We have fabricated a 2D superconducting mesh device by creating a nanohoneycomb array of insulating islands within a LaAlO$_3$/KTaO$_3$ interface superconductor. Magnetoresistance measurements reveal Little-Parks-like quantum oscillations, with a period that closely matches with the threading of a magnetic flux quantum $h/2e$ through the area of a single unit cell of the designed pattern. Notably, these oscillations persist into the anomalous metallic state at the lowest temperatures, providing compelling evidence for Cooper pair transport in this regime. Additionally, we observe an unexpected negative magnetoresistance near zero magnetic field, which can be tuned via $V_G$-controlled $I_G$ and is attributed to strong superconducting fluctuations at the interface. Our findings highlight intriguing transport phenomena arising from the modulation of the superconducting order parameter in periodic interfacial superconducting structures, establishing a promising platform for investigating the origin of the anomalous metallic state and exploring superconductivity under conditions of strong fluctuations.

**Acknowledgments:** We thank H. Yao for helpful discussion. The device was fabricated at the Micro-Nano Fabrication Centre of Zhejiang University. **Funding:** This work was supported by the National Key R&D Program of China (2023YFA1406400), National Natural Science Foundation of China (Grants No. 12325402, 11934016 and 12074334), Innovation Program for Quantum Science and Technology (Grant No.2021ZD0300200), and the Key R&D Program of Zhejiang Province, China (Grants No. 2020C01019 and 2021C01002). **Author contributions:** Y.W. fabricated the device and carried out the experiments under the supervision of Y.X.. S.H. and W.P. contributed to the experimental set-up. Y.W., Y.Z. and Y.X. analyzed the data and wrote the manuscript with inputs from all authors. **Competing interests:** The authors declare no competing interests. **Data and materials availability:** Data for all graphs presented in this paper are available.

# Supplementary Material for

Superconducting quantum oscillations and anomalous negative magnetoresistance in a honeycomb nano-patterned oxide interface superconductor

Yishuai Wang, Siyuan Hong, Wenze Pan, Yi Zhou, and Yanwu Xie*

Correspondence: ywxie@zju.edu.cn

**This file includes:**

>Material and Methods
>Figs. S1-S17
>References

Table of contents

**Materials and Methods.**

1. Device fabrication

The device was fabricated in three steps. Step 1: Patterning the surface of a KTO(110) single crystalline substrate into Hall-bar configuration. Step 2: Depositing LAO film on the patterned substrate. Step 3: Creating an array of insulating islands on the central Hall-bar bridge using cAFM lithography.

**Step 1:** Hall bars were patterned onto the surface of KTO substrate using standard optical lithography and lift off techniques. The hard mask schemed in Fig. 1(a) (see also Fig. S1(a)) was made of a ~175-nm-thick $AlO_x$ film that was fabricated by pulsed laser deposition at room temperature. Note that after the following growth of LAO film, only the Hall-bar area, where the KTO surface is not covered by hard mask, is active; in contrast, the area covered by the hard mask remains highly insulating. To ensure that the whole central Hall-bar bridge can be engineered by cAFM lithography in one run, we chose a small Hall-bar bridge with a size of 10 μm in width and 5 μm in length (the interval between the two voltage contacts).

**Step 2:** The superconducting LAO/KTO interface was formed by depositing a 6-nm thick LAO film using pulsed laser deposition on the KTO substrate that has been patterned as described above. The target of LAO is a single crystal. A 248-nm KrF excimer laser was used. During growth, the temperature was at 300 °C; the atmosphere was a mixture of $1\times10^{-5}$ mbar $O_2$ and $1\times10^{-7}$ mbar $H_2O$ vapor; the laser fluence is ~0.5 $Jcm^{-2}$; the laser repetition rate is 10 Hz. After growth, the sample was cooled down to room temperature under the growth atmosphere.

**Step 3:** The cAFM lithography was carried out using a commercial atomic force microscope machine (Park NX10) at room temperature, under ambient conditions, with a relative humidity of ~45%. Pt/Cr-coated metallic tips (Budget Sensors, Multi75E-G) were used. As shown in Fig. 1(a), the interface insulating islands were produced by briefly touching (40nN, 2s duration) the LAO surface with the charged ($V_{tip}$ = -20V) tip (the interface was grounded). Note that while the cAFM lithography can effectively modify the interface conduction, it does not significantly change the structures of LAO or KTO [26].

2. Surface potential measurements

The surface potential measurements were carried out with single-pass Kelvin probe force microscopy (KPFM) [59] using the same AFM machine (Park NX10) and the same Pt/Cr-coated metallic tips (Budget Sensors, Multi75E-G) at room temperature and under ambient conditions. The measurements were obtained in the non-contact mode without lifting the tip. Using one feedback loop to nullify the electrostatic force and using another feedback loop to maintain topographical scanning, both the topographical image (Fig. S2) and surface potential image (Fig. 1(b)) can be acquired simultaneously [59]. The typical KPFM scan parameters are $V_{ac}$ of 2 V (peak-to-peak), $f_{resonace}$ of 17 kHz, scan rate of 0.2 Hz, and resolution of 256 × 256 pixels.

3. Electrical contacts and transport measurements

**Contacts.** The conducting LAO/KTO interface was contacted by ultrasonic bonding with Al wires.

**Transport measurements.** DC transport measurements were carried out in a commercial $^4$He cryostat with a $^3$He insert (Cryogenic Ltd.). To avoid photoconduction, all transport measurements were performed after keeping the samples in the dark for at least 5 hours. The electrical resistance was measured in a standard four-terminal configuration by using a combination of DC current source and voltmeter (Keithley 6221 and 2182A) with current reversal method. The gate voltage $V_G$ was applied with a source measure unit (Keithley 2450). The data shown in Figs. S15 and S17 were measured using a dilution fridge PPMS (Quantum Design).

**Magnetoresistance.** Magnetic fields were applied perpendicular to the device and ramped with a typical sweep rate of 0.15 mTs$^{-1}$. It should be noted that the raw magnetoresistance data exhibit a slight hysteresis behavior (Fig. S7(a)), which may be related to magnet remanence of the measurement system or trapped magnetic flux in the device [32]. Since hysteresis is not the focus of this study, zero field was reset to the center of the Little-Parks-like oscillations in this paper for a better presentation (Fig. S7(b)).

**Leakage.** As shown in Figs. 1(g) and 1(h), the leakage current $I_G$ was negligible (< 0.1 nA) for $V_G$ between -90 and 60 V, and became gradually and strikingly large with further increasing the absolute value of $V_G$. In this work, the $I_G$ was used as an effective control parameter of the observed ANMR. We point out that although the $I_G$-$V_G$ dependence may vary from sample to sample, for a given sample, as shown in the present device, it is reversible and reproducible after an initial sweeping of $V_G$ between -150 to 150 V.

4. Theoretical calculation

**Theoretical calculation based on classic Little–Parks effect.**

For Little-Parks effect, the amplitude of critical temperature oscillations $\Delta T_c$ are usually given by [31,60–62]:

$$\Delta T_c = 0.14 T_c \left(\frac{\xi_0}{r}\right)^2 \qquad (1)$$

where $T_c$ is the superconducting critical temperature at zero magnetic field, $\xi_0$ is the coherence length, and the loop effective radius $r = (S/\pi)^{1/2}$. The amplitude of $T_c$ oscillations $\Delta T_c$ can be converted to the amplitude of resistance oscillations $\Delta R$ by [31,60–62]:

$$\Delta R = \Delta T_c \frac{dR}{dT} \qquad (2)$$

The zero-temperature Ginzburg–Landau coherence length $\xi_{GL}(0)$ was approximately used as $\xi_0$ which can be extracted using the linearized Ginzburg-Landau form (Eq. (3)), and the differential resistance $dR/dT$ can be obtained from the numerical differentiation of the $R(T)$ curves. The calculated results are shown by the dashed line in Fig. S8(c) and (d).

$$\mu_0 H_{c2}^{\perp}(T) = \frac{\Phi_0}{2\pi\xi_{GL}(0)^2}\left(1 - \frac{T}{T_c}\right) \qquad (3)$$

It should be noted that the Eq. (1) is based on the clean limit case, the dirty limit case is given by [61]:

$$\Delta T_c = 0.18 T_c \frac{\xi_0 l}{r^2} \qquad (4)$$

where $l$ is the electron mean free path. For LAO/KTO(110) interface superconductors, the value of $l$ is about 1/3-1/4 of $\xi_{GL}$ [9], thus the estimated $\Delta R$ is smaller than the clean limit case.

**Theoretical fit based on fluxoid dynamics model.**

In the case of field-driven modulation of the height of the energy barrier to vortex motion, the temperature-dependent amplitude $\Delta R$ can be approximately expressed as [31]:

$$\Delta R = R_N \left(\frac{E_0}{2k_B T}\right)^2 \frac{I_1(\alpha)}{[I_0(\alpha)]^3} \qquad (5)$$

where $\alpha = (E_v + E_0/4)/(2k_B T)$, $I_1$ is the first-order modified Bessel function of the first kind, $R_N$ is the normal-state resistance, $k_B$ is Boltzmann constant, and $T$ is temperature. The energy $E_v$ needed for the creation of the vortex or antivortex in the superconducting ring, and interaction of a vortex or an antivortex with the current associated with the fluxoid in terms of the energy $E_0$ are written as [31,43]:

$$E_v = \frac{\Phi_0^2}{8\pi^2 \Lambda} \ln \frac{2w}{\pi \xi_{GL}} \qquad (6)$$

$$E_0 = \frac{\Phi_0^2}{8\pi^2 \Lambda} \frac{w}{r} \qquad (7)$$

where $\Phi_0$ is the magnetic flux quantum $hc/2e$ (Gaussian units), $\xi_{GL}$ is the Ginzburg–Landau coherence length, $r$ and $w$ are the effective radius and width of the superconducting loop, $\Lambda = 2\lambda^2/d$ is the Pearl penetration depth in a film of thickness $d$ with a penetration depth $\lambda$. For LAO/KTO(110) interface superconductors, the value of the electron mean free path $l$ is much smaller than the Ginzburg–Landau coherence length $\xi_{GL}$ [9]. Therefore, we approximately adopt the dirty limit case for the temperature dependence of coherence length and penetration depth [61,63]:

$$\xi_{GL}(T) = 0.855(\xi_0 l)^{\frac{1}{2}} \left[\frac{T_c}{T_c - T}\right]^{\frac{1}{2}} \qquad (8)$$

$$\lambda(T) = 0.613\lambda_L(0) \left[\frac{\xi_0}{l} \frac{T_c}{T_c - T}\right]^{\frac{1}{2}} \qquad (9)$$

Now consider all the parameters needed to calculate $\Delta R$ through Eq. (5). The effective radius $r = (S/\pi)^{1/2}$, where S is the enclosed area of a hexagonal superconducting ring (Fig. 1(c)). Substituting the onset transition temperature $T_c^{on}$ and resistance $R_N^{on}$ for $T_c$ and $R_N$ (Figs. S8(a)), hence other parameters ($\xi_0$, $\lambda_L(0)$, $l$, $d$ and $w$) can be obtained by fitting to the measured temperature dependence of the amplitude of the Little-Parks-like oscillations. The fit shown by the solid line in Figs. S8(c) and S8(d).

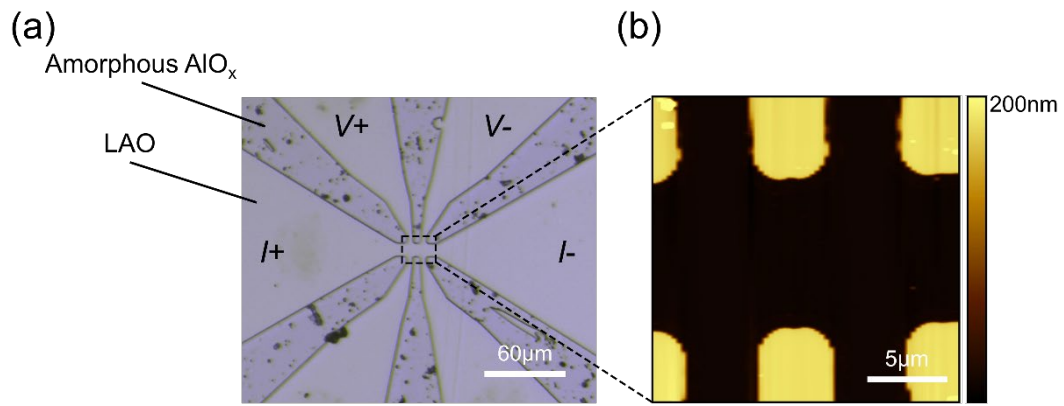

FIG. S1. Photograph (a) and atomic force microscopy image (b) of the LAO/KTO Hall-bar device with a size of 10 μm in width and 5 μm in length (the interval between the two voltage contacts).

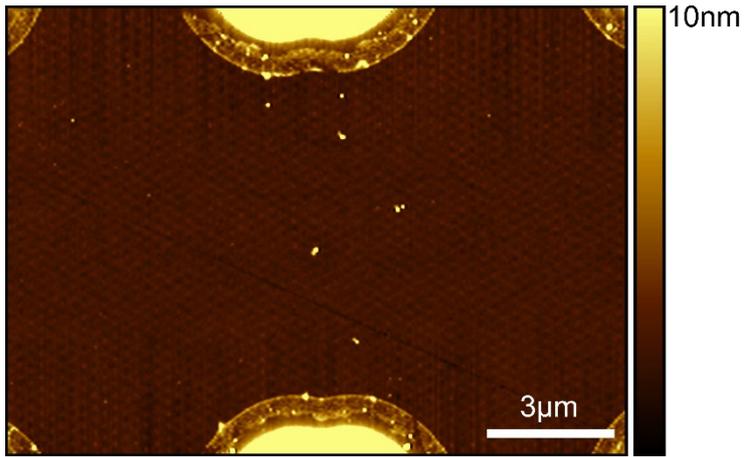

FIG. S2. Surface morphology image of the patterned area as indicated by the dashed white lines in Fig. 1(a).

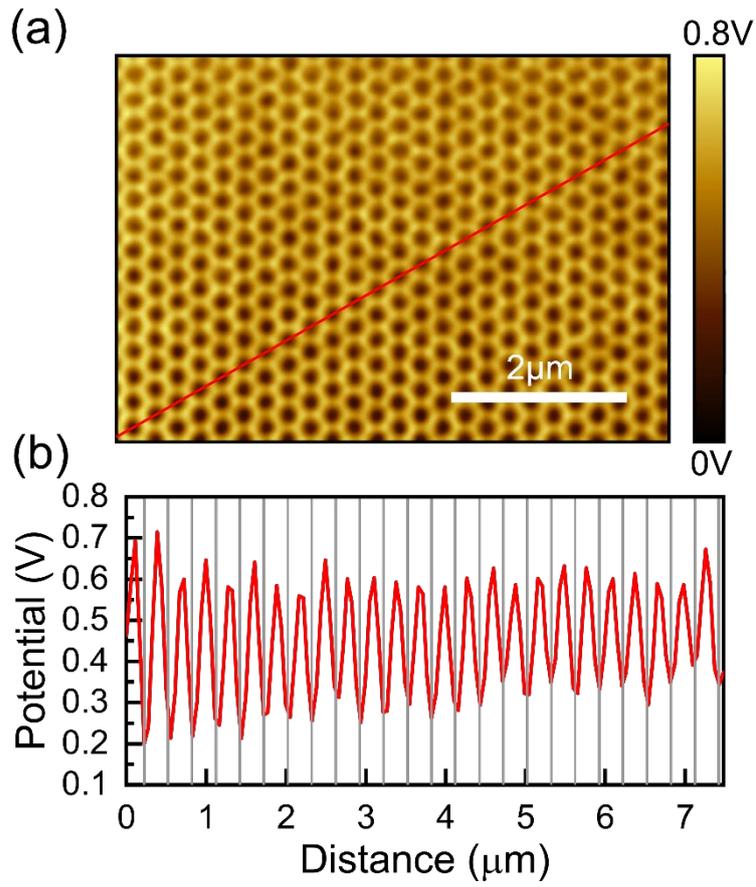

FIG. S3. (a) An enlarged view of a section of Fig. 1(b). (b) Potential profile along the red line in panel (a). The gray grid spacing is 300 nm.

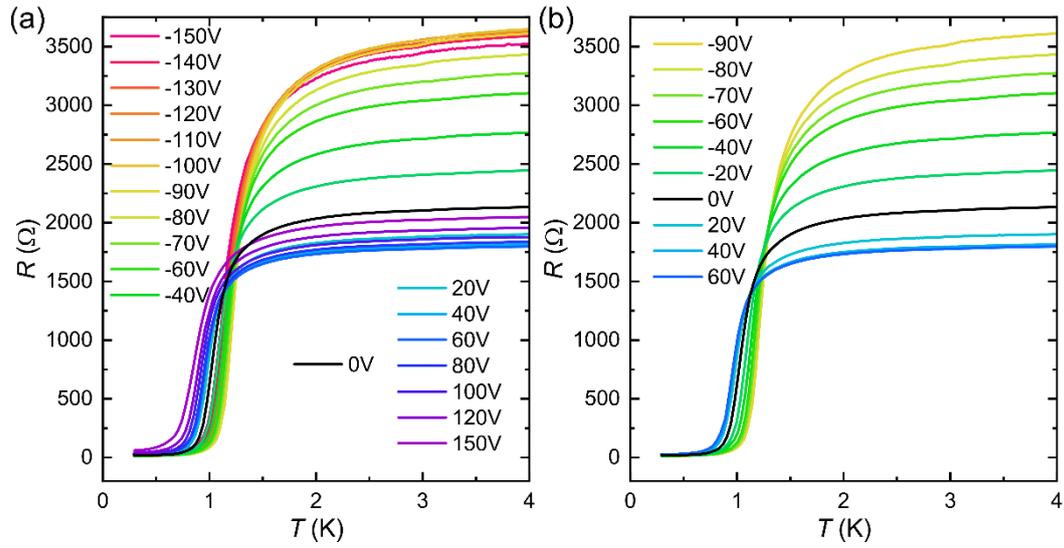

FIG. S4. (a) Full curves of the $V_G$-tuned transport properties of the device. (b) $V_G$-tuned transport properties of the device in the absence of $I_G$.

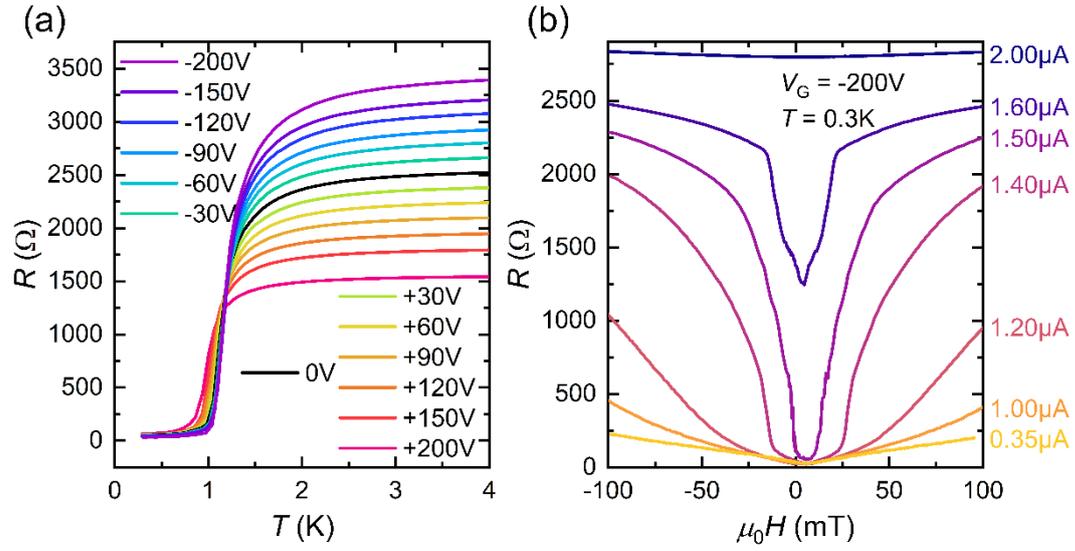

FIG. S5. Transport properties of an LAO/KTO Hall-bar device with a size of 10 μm in width and 5 μm in length. (a) $V_G$-tuned $R(T)$ of the device. (b) $R(H)$ curves measured at different DC currents for $V_G$ = -200 V.

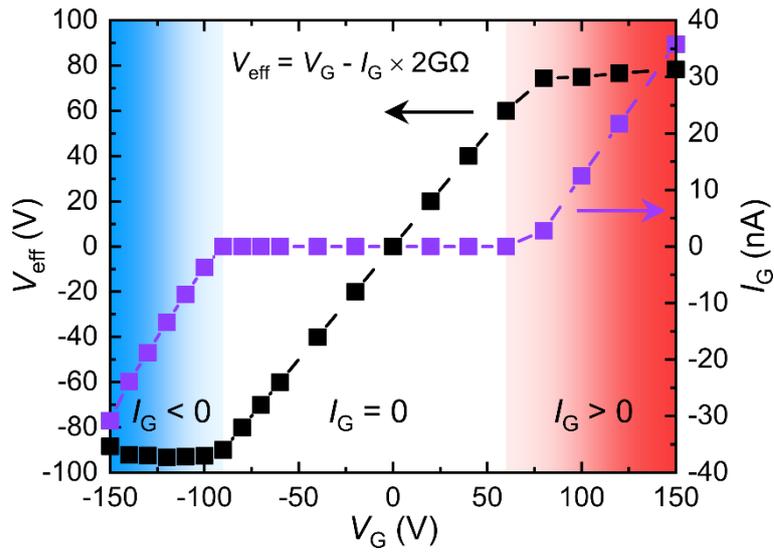

FIG. S6. The effective gate voltage $V_{eff}$ (the voltage between the gate electrode and the conducting interface) as a function of the gate voltage $V_G$.

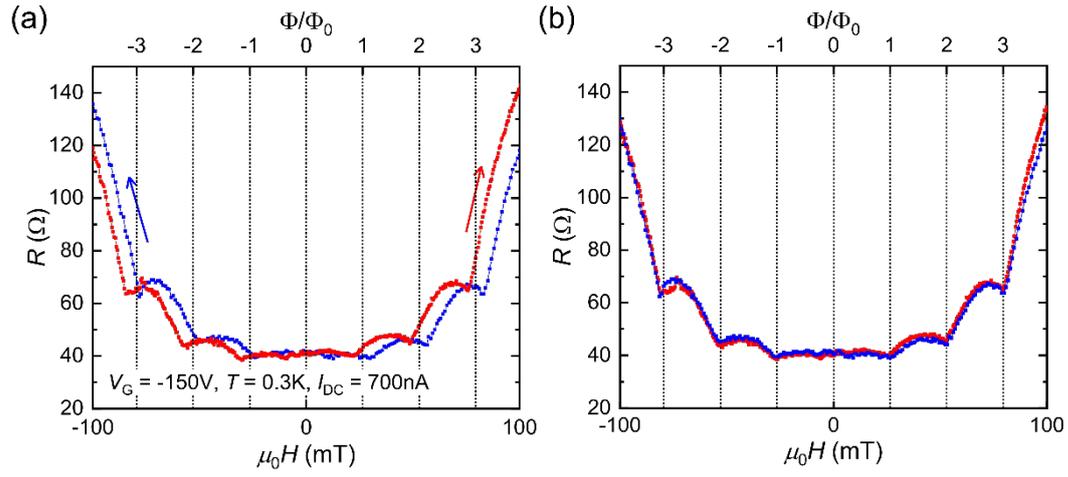

FIG. S7. Centering processing of the magnetoresistance curves. (a) The raw magnetoresistance curves. (b) The magnetoresistance curves after centering. Red curves denote the field sweeping direction from negative to positive; and blue ones denote the reverse sweeping direction.

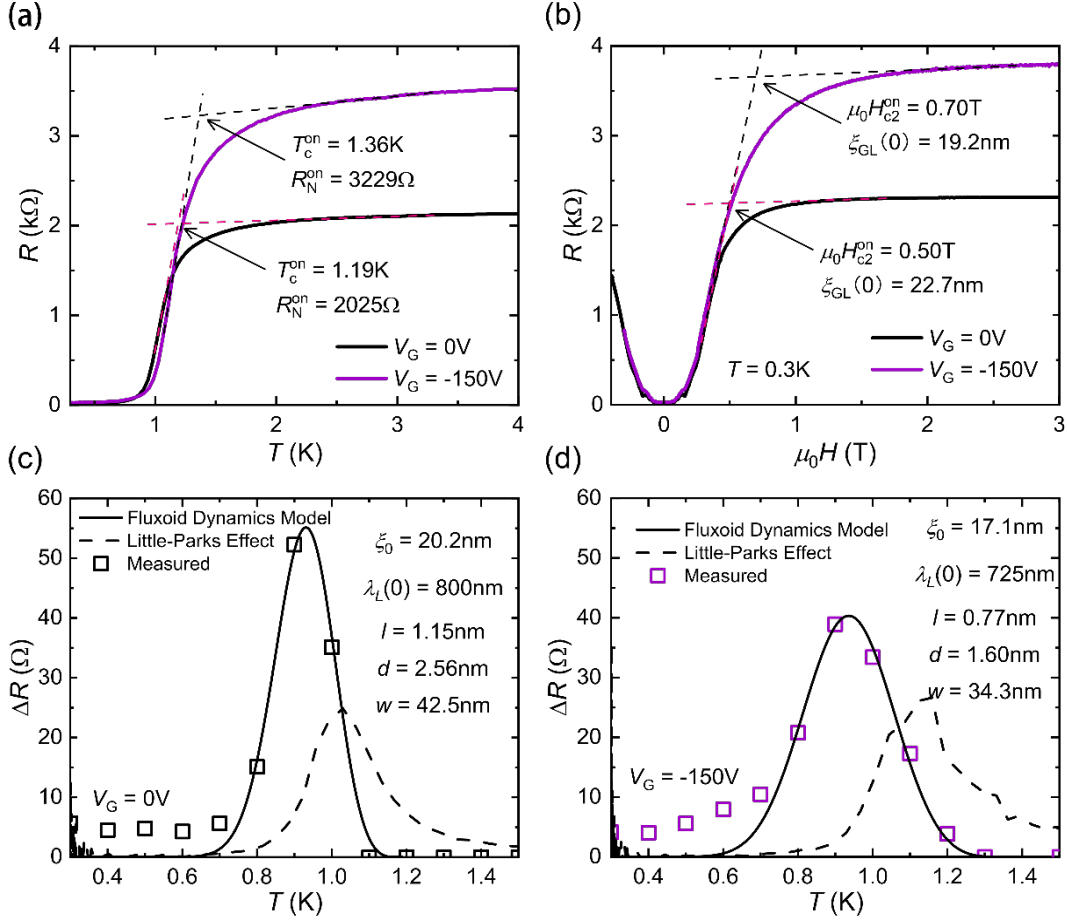

FIG. S8. (a) Temperature dependence of the device resistance in zero magnetic field for $V_G = 0$ and -150 V. (b) Resistance of the device as a function of an applied magnetic field, at $T = 0.3$ K, for $V_G = 0$ and -150 V. The zero-temperature Ginzburg-Landau coherence length $\xi_0$ is extracted using the linearized Ginzburg-Landau form [9] $\mu_0 H_{c2}^{\perp}(T) = [\Phi_0/2\pi\xi_{GL}(0)^2](1 - T/T_c)$, where $\mu_0$ is the vacuum permeability, $\Phi_0$ is the magnetic flux quantum $h/2e$, $T_c$ and $H_{c2}^{\perp}$ is the critical temperature and perpendicular critical magnetic field, the onset transition temperature $T_c^{on}$ and onset transition magnetic field $H_{c2}^{on}$ of 0.3 K were used here. (c), (d) The first oscillation's amplitude of the Little-Parks like oscillations, $\Delta R$ as a function of temperature. The solid lines represent the theoretical fit based on the fluxoid dynamics model, with the fitting parameters displayed in the upper right corner of each figure. The dashed lines are the upper limit for the amplitude of resistance oscillations calculated for the Little–Parks effect.

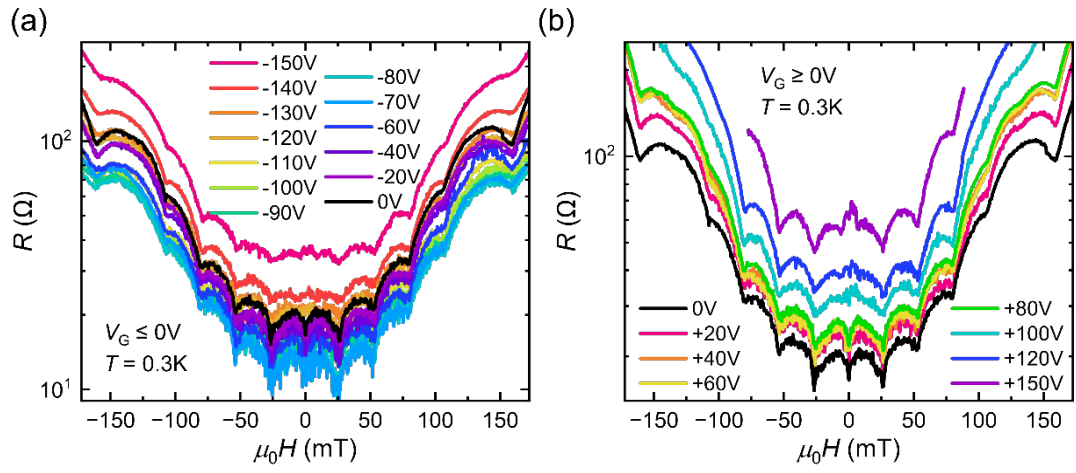

FIG. S9. Magnetoresistance at different $V_G$ values for $I_{DC}$ = 350 nA. (a) $V_G \leq 0$. (b) $V_G \geq 0$.

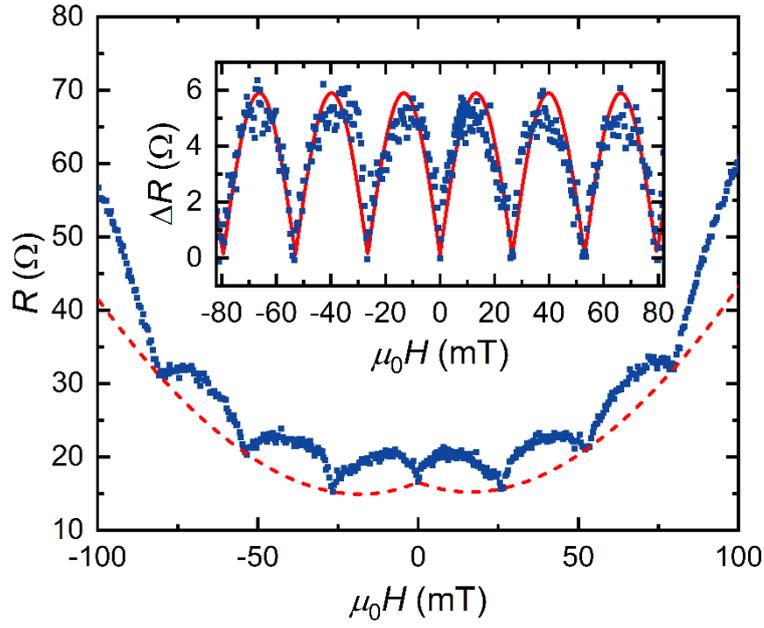

FIG. S10. Replotting the data in Fig. 2(a). The red dashed line denotes the aperiodic background with a weak ANMR. Inset: Oscillation with the background subtracted. The oscillation can be well-fitted by $\Delta R \propto |\sin(\pi\Phi/\Phi_0)|$ (red solid line), which suggests the formation of effective Josephson junctions [51].

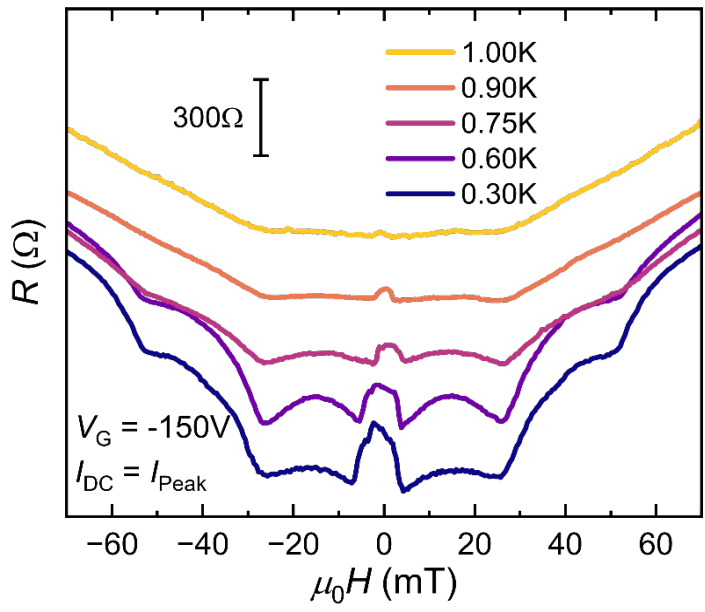

FIG. S11. Magnetoresistance corresponding to Fig. 4(e). Curves are vertically offset for clarity.

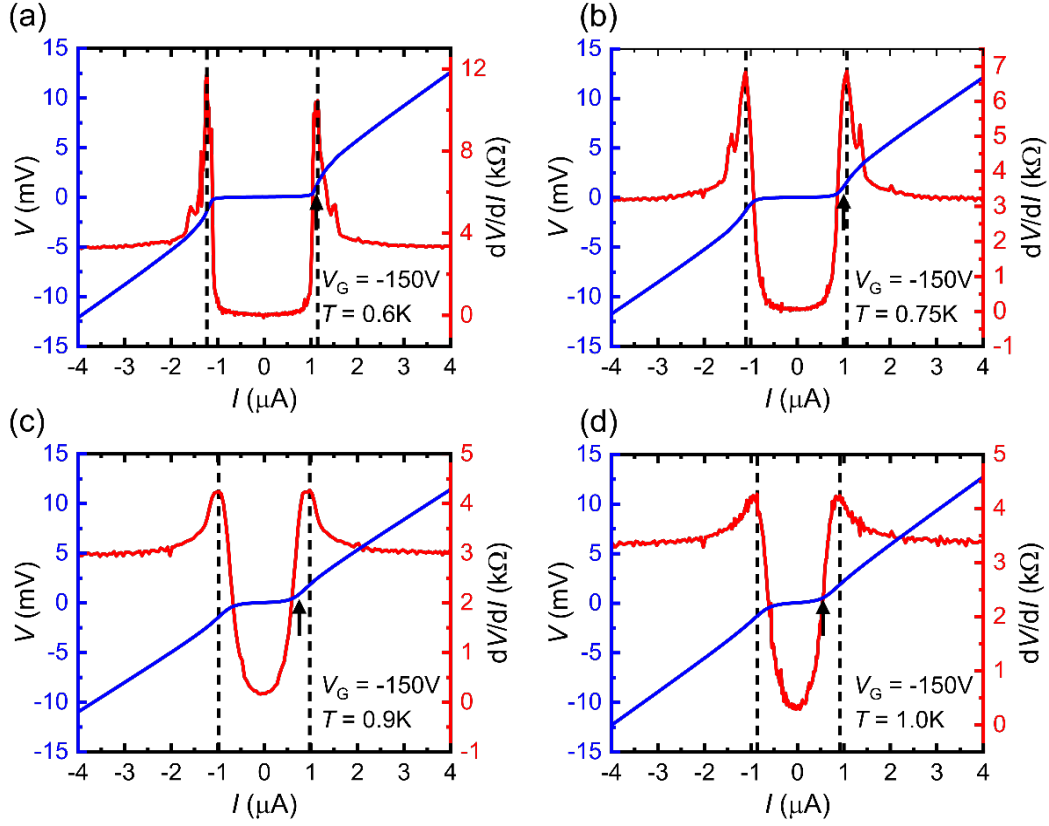

FIG. S12. Temperature-dependent current-voltage (*I-V*) and differential resistance (d*V*/d*I)* curves of the device at $V_G$ = -150 V. The dashed lines indicate the position of the peaks in differential resistance, which are used to define the critical current $I_c$. The arrows mark the position of $I_{peak}$, where the amplitude of the ANMR reaches its maximum. The discrepancy between $I_c$ and $I_{peak}$ can be attributed to the decreasing precision of $I_c$ at higher temperatures, due to the less sharp superconducting transition in the *I-V* curve.

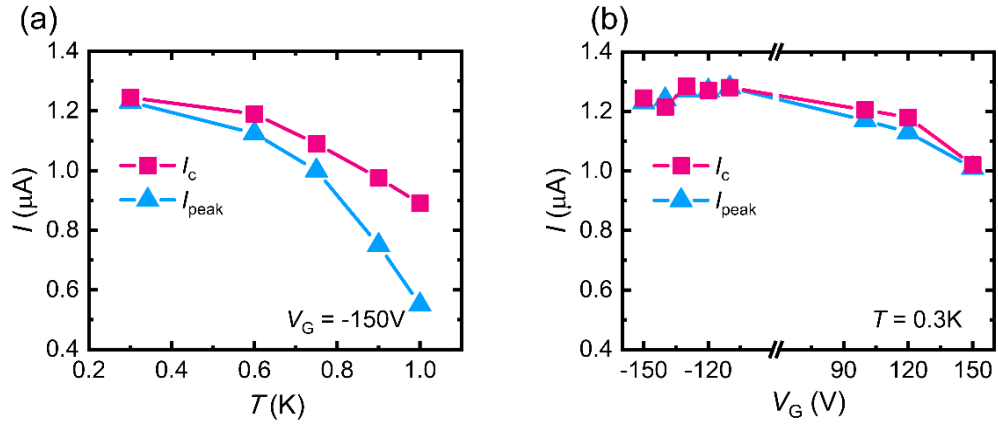

FIG. S13. (a) Temperature dependence of $I_c$ and $I_{peak}$ for the data shown in Figs. 4(d) and 4(f). As discussed in Fig. S12, the discrepancy between $I_c$ and $I_{peak}$ can be attributed to the decreasing precision of $I_c$ at higher temperatures, due to the less sharp superconducting transition in the $I$-$V$ curve. (b) $V_G$ dependence of $I_c$ and $I_{peak}$ for the data shown in Figs. 4(f) and 4(g).

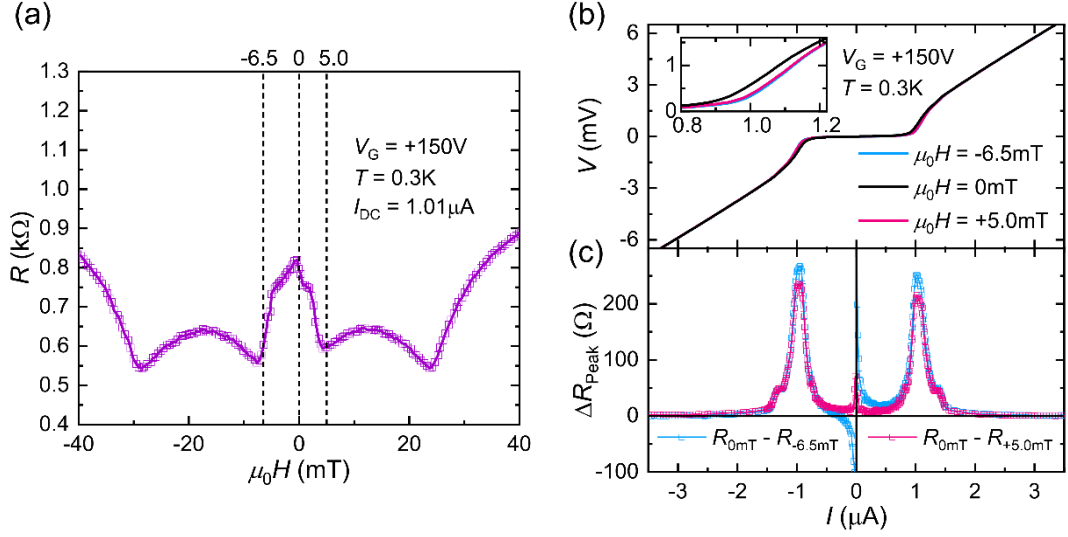

FIG. S14. Negative magnetoresistance induced by the increase in critical current. (a) $R(H)$ curve of the device measured at $V_G$ = +150V, $T$ = 0.3K, and $I_{DC}$ = 1.01μA. (b) $IV$ curves under magnetic fields corresponding to the dashed lines in (a), the inset shows a magnified view around critical current. (c) The difference between the zero-field and low-field resistances, $\Delta R_{Peak}$ as a function of applied current. The values of resistances ($R_{0mT}$, $R_{-6.5mT}$, $R_{+5.0mT}$) were extracted from the $IV$ curves ($R = V / I$).

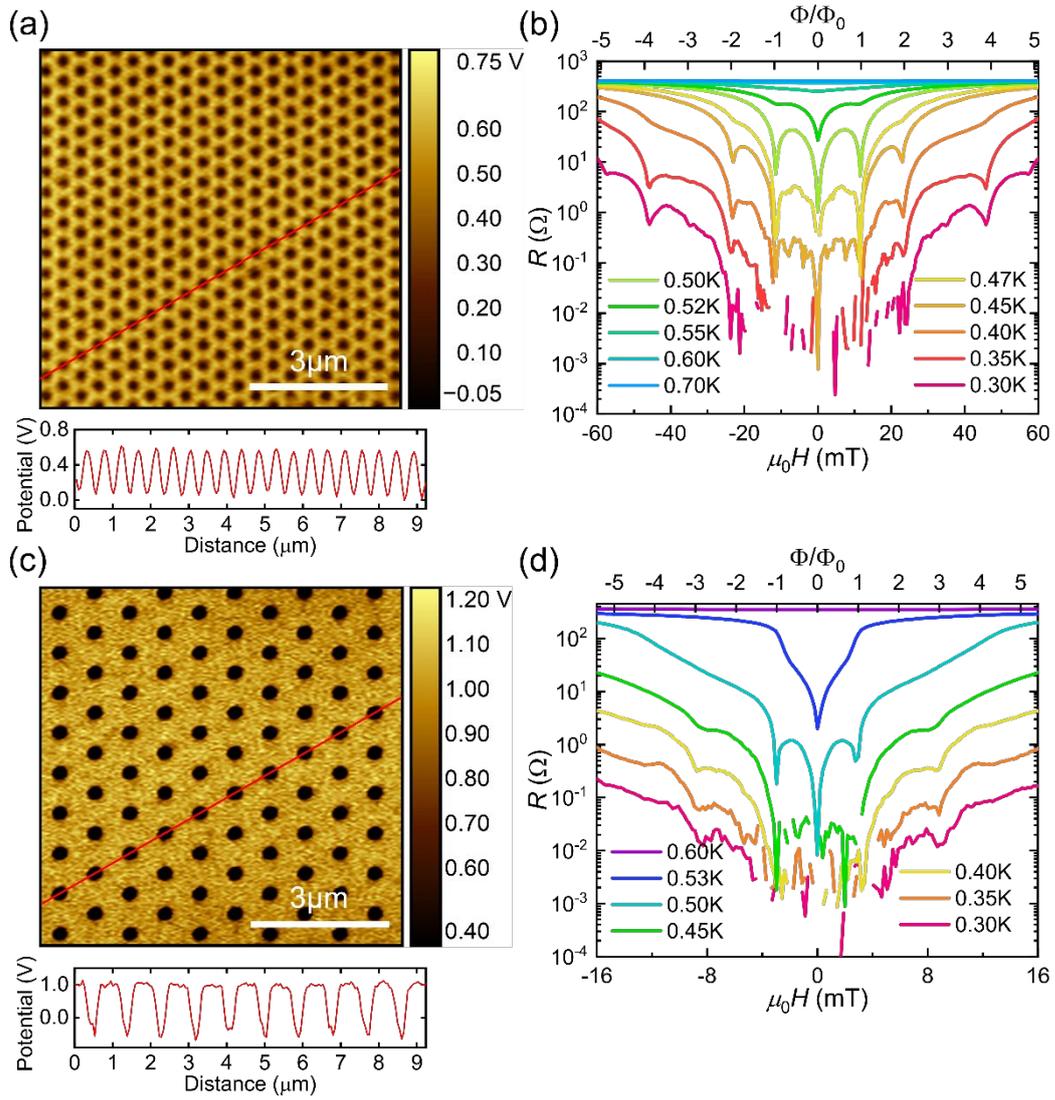

FIG. S15. Surface potential images of devices with a sketched lattice constant of 450 nm (a) and 900 nm (c). The lower panels display the corresponding line profiles indicated in the upper panels. The magnetoresistance at different temperatures of devices with a sketched lattice constant of 450 nm (b) and 900 nm (d).

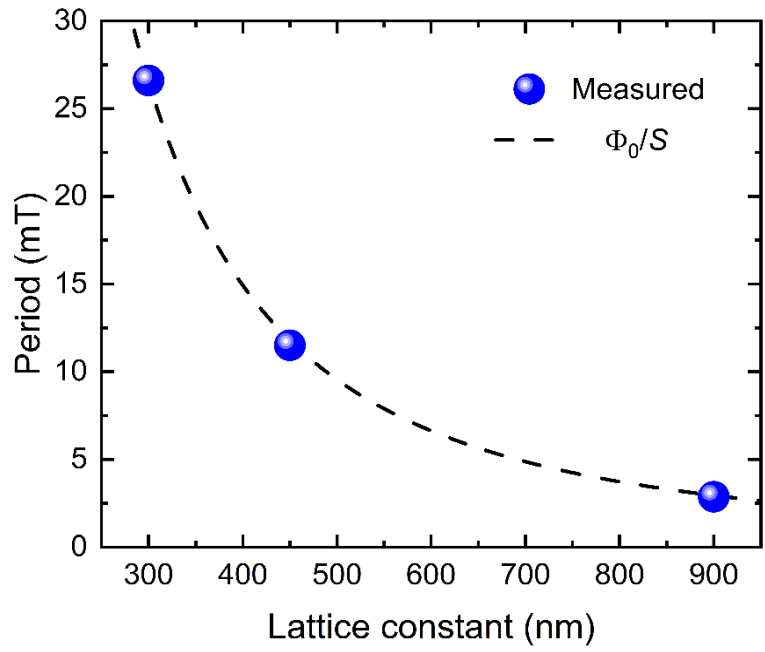

FIG. S16. The period of the Little-Parks-like oscillations as a function of the sketched lattice constant of the nanohoneycomb array.

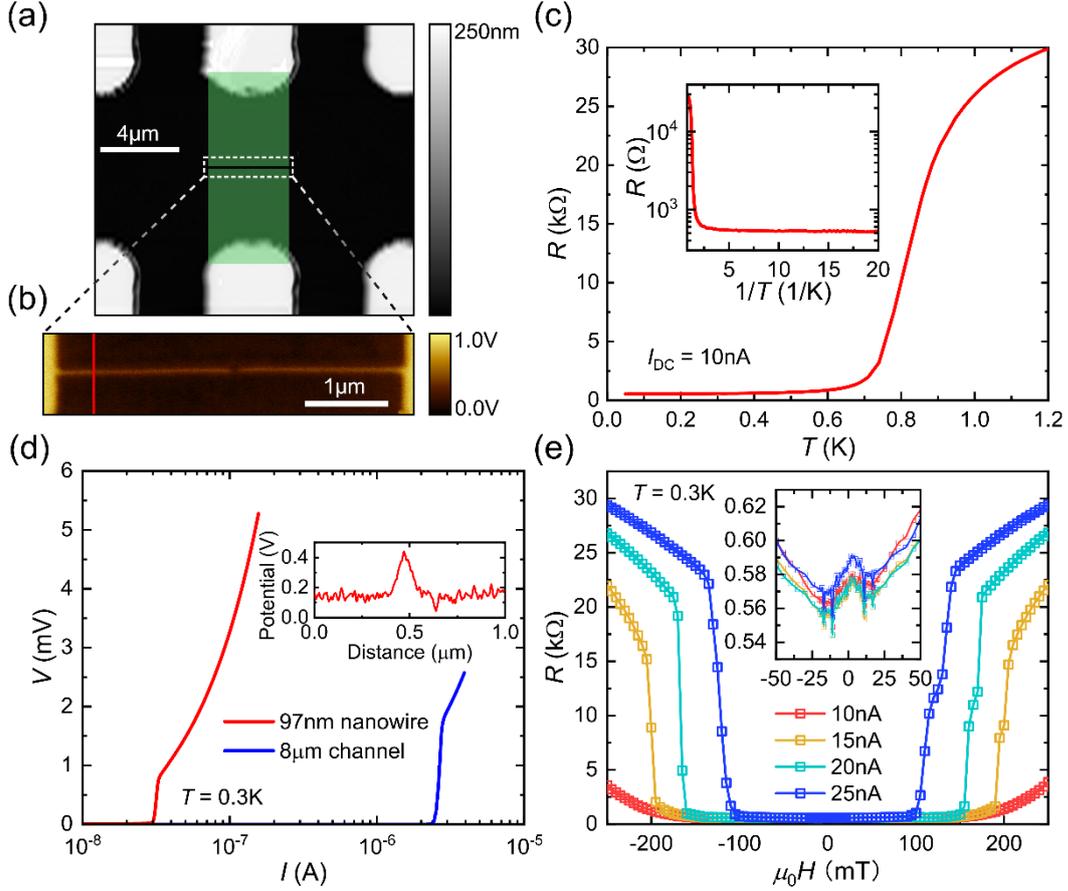

FIG. S17. LAO/KTO(110) nanowire device. (a) Schematic of cAFM lithography pattern overlaid the surface morphology image (Hall-bar device with a size of 8 μm in width and 4 μm in length). The green areas were rendered insulating by cAFM lithography, thus a narrow conducting nanowire was retained in the middle of the Hall channel. (b) Surface potential image of the nanowire as indicated by the dashed white lines in (a). (c) Temperature dependence of the device resistance in zero magnetic field. The inset shows an Arrhenius plot of the $R(T)$ curve. (d) $IV$ curves of the nanowire and the 8-μm-wide Hall-bar channel. The width of the nanowire is estimated to be 97 nm by comparing its critical current with that of the 8 μm channel. Inset: Potential profile across the nanowire along the vertical line indicated in the panel (b). (e) $R(H)$ curves measured at different DC currents. The inset shows a magnified view around zero field.

**Supplementary References**